\documentclass{new_tlp}
\usepackage{mathptmx}
\usepackage{amsmath}
\usepackage{amssymb}
\usepackage{enumerate}

\def\w{\widehat}

\def\rar{\rightarrow}
\def\lrar{\leftrightarrow}

\def\beq{\begin{equation}}
\def\eeq#1{\label{#1}\end{equation}}
\def\ba{\begin{array}}
\def\ea{\end{array}}

\def\proof{\noindent{\bf Proof}.\hspace{3mm}}

\title{\bf Proving Infinitary Formulas} 
\author{AMELIA HARRISON, VLADIMIR LIFSCHITZ\\  
University of Texas, Austin, Texas, USA \\
\email{ameliaj,vl@cs.utexas.edu}\\
\and JULIAN MICHAEL\\
University of Washington, Seattle, Washington, USA\\
\email{julianjohnmichael@gmail.com}}

\begin{document}

\date{}
\maketitle

\begin{abstract}
The infinitary propositional logic of here-and-there is important for the
theory of answer set programming in view of its relation to strongly
equivalent transformations of logic programs.  We know a formal system
axiomatizing this logic exists, but a
proof in that system may include infinitely
many formulas. In this note we describe a relationship between
the validity of infinitary formulas in the logic of here-and-there
and the provability of formulas in some finite deductive systems.
This relationship allows us to use finite proofs to
justify the validity of infinitary formulas.
This note is under consideration for publication in Theory and Practice of Logic 
Programming.  
\end{abstract}

\section{Introduction}\label{sec:intro}


The semantics of ASP programs can be defined using a translation
that turns programs into sets of infinitary propositional formulas 
\cite{geb15}. To prove properties of ASP programs we need then to 
reason about stable models of infinitary formulas in the sense of
Truszczynski \citeyear{tru12}. In particular, we often need to know which 
transformations of infinitary formulas do not affect
their stable models. It is useful to know, for instance, that stable
models of infinitary formulas are not affected by applying the infinitary
De Morgan's laws 
\beq
\bigwedge_{\alpha \in A} \neg F_\alpha \lrar \neg \bigvee_{\alpha \in
A}F_\alpha, 
\eeq{dm2}
and
\beq
\bigvee_{\alpha \in A} \neg F_\alpha \lrar \neg \bigwedge_{\alpha \in
A} F_\alpha
\eeq{dm1}
where $A$ may be infinite.  
``Strongly equivalent'' transformations of 
this kind are used in the proof 
of the interchangeability of
the cardinality constraint $\{p(X)\}0$ and the conditional literal 
$\bot : p(X)$ \nocite{har13c} (Harrison et al.~2015a, Example 7), as
well as the proof  
of correctness of the $n$-queens 
program given in the electronic appendix of \cite{geb15}.  

Strongly equivalent transformations of infinitary formulas are 
characterized by the infinitary logic of here-and-there \cite{har15a}. The set 
of theorems in the sense of that paper coincides with the set of all infinitary 
formulas that are ``HT-valid''---satisfied by all interpretations 
in the sense of the logic of here-and-there. 

The set of theorems is defined by Harrison et al.~\citeyear{har15a}
in terms of closure under a 
set of inference rules; there is no definition of a proof in that paper.
It is possible to reformulate the definition of a theorem in 
terms of proofs, but those proofs would consist generally of 
infinitely many formulas, because some of the inference rules 
introduced there have infinitely many premises. In formalized 
mathematics, proofs are useful in that
they are finite syntactic objects that can establish
the validity of assertions about infinite domains.
``Infinite proofs'', on the other hand, do not have this property.

Can we use finite syntactic objects of some kind to 
establish that an infinitary formula is HT-valid, at least in 
some cases? 

The definition of an instance of a propositional formula 
(Harrison et al.~2015a) may help us answer this question.
Propositions 1 and 3 in that paper show that
substituting infinitary formulas for atoms in a finite 
intuitionistically provable formula results in an HT-valid 
formula. For example, the formula
\beq
(p \lor q) \land r \lrar (p \land r) \lor (q \land r)
\eeq{subsum}
is intuitionistically provable;\footnote{Formalizations of 
propositional intuitionistic logic can be found, for instance, in
Chapters 2 and 8 of Mints's monograph \citeyear{min00}.
Formalizations of 
first-order intuitionistic logic can be found in Chapters 13 and 15
of that book.}
it follows that for any infinitary formulas $F$, $G$, $H$, the infinitary
formula
\beq
(F \lor G) \land H \lrar (F \land H) \lor (G \land H)
\eeq{subsum2}
is HT-valid. We can think of a proof of 
(\ref{subsum}) as a proof of (\ref{subsum2}) with
respect to the substitution that maps $p$ to $F$, $q$ to $G$, and $r$
to $H$.
In a similar way, we can talk about proofs of the formula
\beq
\left (\bigvee_{\alpha \in A} F_\alpha \right ) \land G
\lrar \bigvee_{\alpha \in A} (F_\alpha \land G)
\eeq{subsum3}
for any non-empty
finite family $(F_\alpha)_{\alpha \in A}$ of infinitary formulas  
and any infinitary formula~$G$.

In this paper we show how the idea of an infinitary instance of a
finite formula can be used in a different setting.  We will define
instances for first-order formulas, and that will
allow us, for example, to talk about finite proofs of~(\ref{subsum3}) even
when $A$ is infinite. Consider the signature that has (symbols for) the 
elements of $A$ as object constants, the unary predicate constant 
$P$, and the propositional constant~$Q$. We will see
that (\ref{subsum3}) is the instance of the first-order formula
\beq
\exists x P(x) \land Q \lrar \exists x (P(x) \land Q)
\eeq{subsum4}
corresponding to the substitution that maps $P(\alpha)$ to $F_\alpha$,
and~$Q$ to~$G$.
This formula is intuitionistically provable, and according to the main
theorem of this paper it follows that (\ref{subsum3}) is HT-valid. 

After a review of the infinitary logic of here-and-there in Section 
\ref{sec:def}, we define instances of a first-order formula in Section
\ref{sec:flatsubs}, and state the main theorem in Section \ref{sec:mt}. 
Two other useful forms of the main theorem are discussed in
Section~\ref{sec:so}. The proof of the theorem is outlined in
Section~\ref{sec:proof}. 

A preliminary report on this project was presented at the 8th
Workshop on Answer Set Programming and Other Computing Paradigms
held in Cork, Ireland in 2015.

\section{Infinitary Logic of Here-and-There}\label{sec:def}

This review follows Harrison et al.~(2015, 2015a).

\subsection{Infinitary Formulas}\label{sec:inf_syn}

Throughout this note, we will use $\sigma$ to denote a propositional signature,
that is, a set of propositional atoms.  For every nonnegative integer~$r$,
{\sl (infinitary propositional) formulas (over $\sigma$) of rank~$r$} are
defined recursively, as follows:
\begin{itemize}
\item every atom from~$\sigma$ is a formula of rank~0;
\item if $\mathcal{H}$ is a set of formulas, and~$r$ is the smallest
nonnegative 
integer that is greater than the ranks of all elements of $\mathcal{H}$,
then $\mathcal{H}^\land$ and $\mathcal{H}^\lor$ are formulas of rank~$r$;
\item if $F$ and $G$ are formulas, and~$r$ is the smallest nonnegative
integer that is greater than the ranks of~$F$ and~$G$, then $F\rar G$ is a
formula of rank~$r$.
\end{itemize}
We will write $\{F,G\}^\land$ as $F\land G$, and
$\{F,G\}^\lor$ as $F\lor G$.
The symbols $\top$ and $\bot$ will be understood as abbreviations
for~$\emptyset^{\land}$ and   
for~$\emptyset^{\lor}$ respectively; $\neg F$ and $F \lrar G$ are understood
as abbreviations in the usual way. 

A set or family of formulas is {\sl bounded} if the ranks of its members
are bounded from above.  For any bounded family $(F_\alpha)_{\alpha \in A}$
of formulas, we denote the formula 
$\{F_\alpha: {\alpha \in A}\}^\land$  by $\bigwedge_{\alpha \in A} F_\alpha$, 
and similarly for disjunctions. 
For example, if all formulas $F_\alpha$ and $G$ are atoms then the
left-hand side of equivalence~(\ref{subsum3}) is shorthand for the
formula
$$\left\{\left\{F_\alpha : \alpha\in A\right\}^\lor,G\right\}^\land$$
of rank 2.

\subsection{HT-Interpretations}

An {\sl HT-interpretation} of~$\sigma$ is an ordered pair
$\langle I^h,I^t\rangle$ of subsets of~$\sigma$ such that
$I^h\subseteq I^t$. 
The symbols $h, t$ are called {\sl worlds}; respectively {\sl here} and {\sl there}.
They are ordered by the relation $h < t$. HT-interpretations are the special
case of Kripke models for intuitionistic 
logic\footnote{\tt http://plato.stanford.edu/entries/logic-intuitionistic/\#KriSemForIntLog} 
with only two worlds. 

The satisfaction relation between an HT-interpretation $I = \langle I^h, I^t
\rangle$, a world~$w$, and a 
formula is defined recursively, as follows:
\begin{itemize}
\item $I, w\models p$ if $p\in I^w$;
\item $I, w\models\mathcal{H}^\land$ if for every formula
$F$ in~$\mathcal{H}$, $I, w\models F$;
\item $I, w\models\mathcal{H}^\lor$ if there is a formula
$F$ in~$\mathcal{H}$ such that $I, w\models F$;
\item
$I, w\models F\rar G$ if, for every world $w'$ such that $w \leq w'$, 
$\;\;I, w'\not \models F$ or $I, w' \models G$. 
\end{itemize}
In particular,

\medskip
$\quad \;I, w\models \neg F$ if, for every world $w'$ such that $w \leq w'$, 
$\;\;I, w'\not \models F$.

\medskip
We say that~$I$ {\sl satisfies} $F$, and write $I \models F$,
if $I, h \models F$ (equivalently, if
$I,w \models F$ for every world~$w$).  A formula is {\sl HT-valid} 
if it is satisfied by all HT-interpretations.

\section{Substitutions and Instances}\label{sec:flatsubs}

By $\Sigma$ we denote an arbitrary signature in the sense of first-order
logic  that contains at least one object constant. The signature may
include propositional constants (viewed as predicate constants of arity
0). Object constants will be
viewed as function constants of arity 0. In first-order formulas over 
$\Sigma$, we treat the binary connectives $\land$, $\lor$, and $\rar$ and
the  0-place connective~$\bot$ as primitive; $\top$, $\neg$, and $\lrar$
are the usual abbreviations from propositional logic.

A {\sl substitution} is a function $\psi$ that maps 
each closed atomic formula 
over $\Sigma$ to an infinitary formula over $\sigma$, such that the
range of~$\psi$ is bounded. 
A  substitution~$\psi$ is extended from closed atomic formulas to arbitrary 
closed first-order formulas over~$\Sigma$ as follows:

\begin{itemize}

\item $\psi \bot$ is $\bot$;

\item $\psi(\alpha_1 = \alpha_2)$, where $\alpha_1, \alpha_2$ are ground terms, is 
$\top$ if $\alpha_1$ is $\alpha_2$, and $\bot$ otherwise;

\item $\psi(F \odot G)$, where $\odot$ is a binary connective, is  
$\psi F \odot \psi G$; 

\item $\psi \forall v F$  is $\bigwedge_{\alpha} \psi F^v_{\alpha}$,
where $\alpha$ ranges over the ground terms of $\Sigma$;\footnote{By
$F^v_{\alpha}$ we denote the result of substituting $\alpha$ for all free
occurrences of~$v$ in~$F$.}

\item $\psi \exists v F$  is $\bigvee_{\alpha} \psi F^v_{\alpha}$, 
where $\alpha$ ranges over the ground terms of $\Sigma$. 

\end{itemize}
The formula $\psi F$ will be called the {\sl instance of $F$ with respect
to $\psi$}.  

For example, if $\Sigma$ includes the elements of $A$ as object 
constants, but no other function constants,
then~(\ref{subsum3}) is the instance of (\ref{subsum4})
with respect to the substitution $\psi$ defined as follows:
$$
\ba l
\psi P(\alpha) = F_\alpha,\\
\psi Q = G.
\ea
$$
If the function constants of $\Sigma$ are the object constant $a$ and the
unary function constant $s$, then any infinite conjunction of the form
$$
\bigwedge_{i \geq 0} (F_i \rar G_i),
$$ 
where $F_i, G_i$ are infinitary formulas, is the instance
of the first-order formula
$$
\forall x (P(x) \rar Q(x))
$$
with respect to the substitution $\psi$ defined as follows:
$$
\psi(P(s^i(a))) = F_i,
$$
\vskip -6mm
$$
\psi(Q(s^i(a))) = G_i.
$$

\section{Main Theorem}\label{sec:mt}

The main theorem stated below shows that if a closed first-order formula
is intuitionistically provable then all its instances are HT-valid. The
theorem is actually more general because it refers to a deductive system
that includes, in addition to the axioms and inference rules of first-order
intuitionistic logic with equality, some additional axioms. 
We can add, first of all, the axiom schema\beq
F \lor (F \rar G) \lor \neg G 
\eeq{hosoi}
\cite{hos66,ume59}, the axiom schema
\beq
\exists x(F \rar \forall x F)
\eeq{sqht}
\cite{lif07a}, and the 
``decidable equality'' axiom
\beq
x = y \lor x \not = y. 
\eeq{de}
We include also the axioms of the Clark Equality Theory \cite{cla78}: 
\beq
f(x_1, \dots, x_n) \not = g(y_1, \dots, y_m)
\eeq{una}
for all pairs of distinct function constants $f$, $g$ from~$\Sigma$; 
\beq
f(x_1, \dots, x_n) = f(y_1, \dots, y_n)  \rar 
(x_1 = y_1 \land \dots \land x_n = y_n)
\eeq{fc11}
for all function constants $f$ from $\Sigma$ of arity greater than $0$; and 
\beq
t(x) \not = x
\eeq{cet3}
for all terms $t(x)$ that contain $x$ but are different from $x$. 

The deductive system obtained from first-order intuitionistic logic with 
equality by adding axioms (\ref{hosoi})--(\ref{cet3}) will be
denoted by $\mathbf{HHT}$ (``Herbrand logic of here-and-there'').  

\medskip \noindent {\bf Main Theorem. } {\sl If a closed first-order formula $F$ 
is provable in $\mathbf{HHT}$ then any instance of $F$ is 
HT-valid.}

\medskip
\noindent {\bf Example 1. } The infinitary De Morgan's laws (\ref{dm2})
and (\ref{dm1}) with non-empty $A$ are HT-valid because they are 
instances of the first-order formulas
$$
\forall x \neg P(x) \lrar \neg \exists x P(x)
$$
and
$$
\exists x \neg P(x) \lrar \neg \forall x P(x)
$$  
respectively, and these formulas are provable in $\mathbf{HHT}$. (The
first equivalence, and one direction of the second, are provable
intuitionistically.  To prove the second equivalence right-to-left,
use~(\ref{sqht}) with $P(x)$ as $F$.)

If $A$ is empty then formula \eqref{dm2} is $\top \lrar \neg \bot$ and
\eqref{dm1} is $\bot \lrar \neg \top$. Both of these formulas are 
HT-valid. However, in view of the restriction that $\Sigma$ contain at least 
one object constant neither is an instance of the formulas in 
the previous example. Without that restriction, the assertion of 
the Main Theorem would become incorrect. Indeed, the formula $\top \rar 
\bot$ would be then an instance of the intuitionistically provable formula $\forall 
x \; P(x)  \rar \exists x \; P(x)$.  

\medskip
\noindent {\bf Example 2. } As discussed above, the fact that
formula~(\ref{subsum3}) is HT-valid follows from the provability
of~(\ref{subsum4}) in first-order intuitionistic logic.  Consider the
formula dual to~(\ref{subsum3}):
$$
\left (\bigwedge_{\alpha \in A} F_\alpha \right ) \lor G
\lrar \bigwedge_{\alpha \in A} (F_\alpha \lor G).
$$
(As before, $(F_\alpha)_{\alpha \in A}$ is a non-empty family of
infinitary formulas, and $G$ is an infinitary formula.)  The fact that
this formula is HT-valid can be
derived from the main theorem above in a similar way, with
the corresponding first-order formula
$$\forall x P(x) \lor Q \lrar \forall x (P(x) \lor Q).$$
The proof of the right-to-left direction will use~(\ref{sqht}), again
with $P(x)$ as $F$.

\medskip
\noindent {\bf Example 3. } Any formula of the form
$$
\left ( \left ( \bigvee_{\alpha \in A}F_\alpha\right ) \rar G \right ) \lrar
\bigwedge_{\alpha \in A}(F_\alpha\rar G)
$$
with non-empty $A$ (Harrison et al.~2015a,~Example 2)
is HT-valid because it is an instance of the 
intuitionistically provable formula
$$
\left ( \exists x P(x) \rar Q \right ) \lrar \forall x (P(x) \rar Q).
$$

\medskip

\noindent {\bf Example 4. } Any formula of the form 
$$
\bigvee_{\alpha \in A} \left ( F_\alpha \rar \bigwedge_{\beta \in A} F_\beta 
\right ),
$$
where $A$ is non-empty, is HT-valid because it is an instance of the axiom
schema~(\ref{sqht}).  

\section{Including Restrictors}\label{sec:r}

Under the definition of an instance above,
all infinitary conjunctions and disjunctions in an instance of a
formula have the same indexing set.
In this section we give a more general definition that overcomes this
limitation.

We assume here that some unary predicate symbols of the
signature~$\Sigma$ may be designated as {\sl restrictors}.   The role of
restrictors will be somewhat similar to the role of sorts in a many-sorted
signature.  A {\sl generalized variable} is defined as either a variable
or an expression of the form
\beq
(x_1\!:\!R_1,\dots,x_n\!:\!R_n)
\eeq{gv}
where $x_1,\dots,x_n$ ($n\geq 1$) are distinct variables, and $R_1,\dots,R_n$
are restrictors.  {\sl Formulas
with restrictors} are defined recursively in the same way as first-order
formulas over $\Sigma$ except that a quantifier may be followed 
by a generalized variable.  For instance, if
$\Sigma$ includes the unary predicate constants $P$ and $R$, and the
latter is a restrictor, then
\beq
\forall x P(x) \rar \forall (x\!:\!R) P(x)
\eeq{r0}
is a formula with restrictors.

Generalized variables~(\ref{gv}) can be eliminated from a formula with
restrictors by replacing subformulas of the form
$$\forall (x_1\!:\!R_1,\dots,x_n\!:\!R_n) F$$
with
$$\forall x_1\dots x_n(R_1(x_1) \land\cdots\land R_n(x_n) \rar F),$$
and subformulas of the form
$$\exists (x_1\!:\!R_1,\dots,x_n\!:\!R_n) F$$
with
$$\exists x_1\dots x_n(R_1(x_1) \land\cdots\land R_n(x_n) \land F).$$
To prove a formula with restrictors in a deductive system means to prove the
first-order
formula obtained by this transformation. For instance, we
can say that formula~(\ref{r0}) is provable in the intuitionistic
predicate calculus because the formula
$$\forall xP(x) \rar \forall x(R(x)\rar P(x))$$
is provable in that deductive system.
Satisfaction of closed formulas with restrictors is defined in a similar
way.

In the presence of restrictors, a {\sl substitution} is
defined as a function $\psi$ that maps each closed atomic formula~$F$
over $\Sigma$ to one of the formulas $\top$, $\bot$, if $F$ begins with a
restrictor, and to an infinitary formula over $\sigma$ otherwise, such
that the range of~$\psi$ is bounded.  A substitution~$\psi$ is extended
to closed first-order formulas over~$\Sigma$
with restrictors in the same way as for first-order formulas as in
Section~\ref{sec:flatsubs}, with the additional clauses:

\begin{itemize}
\item $\psi\,\forall (x_1\!:\!R_1,\dots,x_n\!:\!R_n) F$ is
 $$ \bigwedge_{\alpha_1,\dots,\alpha_n\ :
               \ \psi R_1(\alpha_1) = \dots =\psi R_n(\alpha_n)=\top}
               \psi F^{x_1 \cdots x_n}_{\alpha_1 \cdots \alpha_n}, $$

\item $\psi\,\exists (x_1\!:\!R_1,\dots,x_n\!:\!R_n) F$ is
       $$ \bigvee_{\alpha_1,\dots,\alpha_n\ :
               \ \psi R_1(\alpha_1) = \dots =\psi R_n(\alpha_n)=\top}
               \psi F^{x_1 \cdots x_n}_{\alpha_i \cdots \alpha_n}. $$
 \end{itemize}

\medskip \noindent {\bf Main Theorem for Formulas with Restrictors.}
{\sl If a closed first-order formula $F$ with restrictors is
provable in $\mathbf{HHT}$ then any instance of $F$ is HT-valid.}

\medskip\noindent {\bf Example 5. }
Consider a formula of the form
\beq
\bigwedge_{\alpha\in A} F_\alpha \rar \bigwedge_{\alpha\in B} F_\alpha,
\eeq{x5}
where~$B$ is a proper subset of $A$.  It
is an instance of~(\ref{r0}): take the elements of $A$ to be the only
function constants of $\Sigma$, and define the substitution~$\psi$ by the
conditions
$$
\ba l
\psi R(\alpha) = \top\hbox{ iff }\alpha\in B,\\
\psi P(\alpha) = F_\alpha.
\ea
$$
Since~(\ref{r0}) is intuitionistically provable,
(\ref{x5}) is HT-valid.

\noindent {\bf Example 6. } 
Any formula of the form
\beq
\bigvee_{\alpha \in A} F_\alpha \land \bigvee_{\beta \in B} G_\beta
                    \lrar
\bigvee_{(\alpha, \beta) \in A \times B} (F_\alpha \land G_\beta)
\eeq{eq:disj-dist}
is an instance of the formula
\beq
\exists(x\!:\!R_1) P(x) \land \exists(y\!:\!R_2) Q(y)
                    \lrar
\exists(x\!:\!R_1,y\!:\!R_2) (P(x) \land Q(y)).
\eeq{r1}
Indeed, we can include the elements of $A\cup B$ among
the object constants of~$\sigma$ and choose $\psi$ so that
$$
\ba l
\psi R_1(\alpha)=\top\hbox{ iff }\alpha\in A,\\
\psi R_2(\alpha)=\top\hbox{ iff }\alpha\in B,\\
\psi P(\alpha)=F_\alpha\hbox{ for all }\alpha\in A,\\
\psi Q(\alpha)=G_\alpha\hbox{ for all }\alpha\in B.
\ea
$$
Since~(\ref{r1}) is intuitionistically provable,
(\ref{eq:disj-dist}) is HT-valid.

References to the new version of the main theorem can be replaced
in some cases by references to the more restricted version from
Section~\ref{sec:mt} at the cost of using more complicated substitutions.
For instance, the claim that formula~(\ref{x5}) is HT-valid, under the
additional assumption that~$B$ is non-empty, can be justified as follows.
Take~$\Sigma$ to be the signature consisting of the elements of~$A$ as
object constants, the unary function constant~$f$, and the unary
predicate constant~$P$.  Choose an element~$\alpha_0$ of~$B$.
Then~(\ref{x5}) is the instance of the formula
$$\forall xP(x) \rar \forall xP(f(x))$$
with respect to the substitution~$\psi$ defined by the condition: for
all object constants~$\alpha$,
\begin{align*}
&\psi P(\alpha)=F_\alpha,&&\\
&\psi P(f^i(\alpha))=F_\alpha &&\hbox{ if $i\geq 1$ and $\alpha\in B$},\\
&\psi P(f^i(\alpha))=F_{\alpha_0} &&\hbox{ if $i\geq 1$ and $\alpha\not\in B$}.
\end{align*}

\section{Including Second-Order Axioms}\label{sec:so}

We will define now an extension $\mathbf{HHT}^2$ of $\mathbf{HHT}$ where 
predicate and function variables of arbitrary arity are included in the 
language, as in Section~1.2.3 of the handbook chapter by Lifschitz et
al.~\citeyear{lif08b}. The set of axioms and inference rules of 
$\mathbf{HHT}$ is extended by adding the 
usual postulates for second-order quantifiers, the axiom  
schema of comprehension
\beq
\exists p \forall x_1 \dots x_n(p(x_1, \dots, x_n) \lrar F) 
\eeq{co}
$(n \geq 0)$, where the predicate variable $p$ 
is not free in $F$, and the axiom of choice
\begin{align}
\label{ch}
\begin{split}
\forall x_1 \dots x_n & \exists x_{n+1}\; p(x_1, \dots, x_{n+1}) \rar \\ & \exists f 
\forall x_1 \dots x_n(p(x_1, \dots, x_n, f(x_1, \dots, x_n)))
\end{split}
\end{align}
$(n > 0)$. The main theorem can be extended as follows. 

\medskip \noindent {\bf Main Theorem for $\mathbf{HHT^2}$. } {\sl If a closed 
first-order formula $F$ (possibly with 
restrictors) is provable in $\mathbf{HHT^2}$ then any 
instance of $F$ is HT-valid.}

\medskip

In the special case when the signature $\Sigma$ contains finitely many function 
constants, by DCA we denote the domain closure axiom: 
$$
\forall p \left ( \bigwedge C_f(p) \; \rar \; \forall x \; p(x) \right )
$$
where the conjunction extends over all function constants $f$ from $\Sigma$, and 
$C_f(p)$ (``set $p$ is closed under $f$'') stands for the formula
$$
\forall x_1 \dots x_n ( p(x_1) \land \dots \land p(x_n) \rar p(f(x_1, \dots,
x_n)).
$$
(In the presence of DCA, axioms (\ref{de}) and (\ref{cet3}) become redundant.)
For instance, if $\Sigma$ contains an object constant $a$ and unary 
function constant $s$ and no other function constants, then DCA  turns into 
the second-order axiom of induction
\beq
\forall p \left ( p(a) \land \forall x \left (p(x) \rar p\left (s(x)
\right ) \right ) \rar \forall x \; p(x) \right ),
\eeq{axoi}
and $\mathbf{HHT}^2 +$ DCA becomes an extension of second-order intuitionistic 
arithmetic. 

In the following version of the main theorem, the signature $\Sigma$ is assumed 
to contain finitely many function constants. 

\medskip \noindent {\bf Main Theorem for $\mathbf{HHT}^2 +$ DCA. } 
{\sl If a closed first-order formula $F$ (possibly with restrictors) is provable in $\mathbf{HHT}^2 +$  DCA
then any  instance of $F$ is HT-valid.}

\medskip

Note that both versions of the main theorem stated in this section refer to
first-order formulas provable using second-order axioms. The notion
of a substitution is not defined here for second-order formulas.

\medskip

\noindent {\bf Example 7. } 
Any equivalence of the form
$$
\left ( F_0\land\bigwedge_{i\geq 0}(F_i\rar F_{i+1}) \right ) \lrar
\bigwedge_{i\geq 0}F_i$$  
(Harrison et al.~2015a, Example 1) is HT-valid. Indeed, with the appropriate 
choice of the signature $\Sigma$, it is an instance of the 
formula
$$
 P(a) \land \forall x (P(x) \rar P(s(x))) \lrar \forall x  P(x).
$$
This formula is provable in $\mathbf{HHT}^2 +$ DCA. 
(The implication left-to-right is given by axiom (\ref{axoi}).)

\section{Proof of Main Theorem}\label{sec:proof}

The proof of the theorem makes use of ``Herbrand HT-interpretations''---Kripke 
models with two worlds and with 
the universe consisting of all ground terms of the signature $\Sigma$.
We will see that all 
theorems of $\mathbf{HHT}$ (and its extensions discussed in the previous 
section) are satisfied by all Herbrand HT-interpretations. 
On the other hand, for any substitution $\psi$ and any HT-interpre\-tation $I$ 
of $\sigma$, we can find 
an Herbrand HT-interpretation $J$ such that $J$ satisfies a closed first-order formula~$F$
if and only if $I$ satisfies $\psi F$. The main theorem will directly follow 
from these two facts. 
 
An {\sl Herbrand HT-interpretation} of a first-order signature
$\Sigma$ is a pair $\langle J^h, J^t\rangle$ of subsets of the Herbrand base of 
$\Sigma$ (that is, the set of all ground atomic formulas over $\Sigma$
that do not include equality)
such that $J^h \subseteq J^t$. By $\mathcal{U}$ we 
denote the Herbrand universe of $\Sigma$, that is, the set of all ground terms 
over $\Sigma$.  

For each function $\mathfrak{f}$ of arity~$n > 0$ that maps from $\mathcal{U}^n$ 
to $\mathcal{U}$ we introduce a function constant $\mathfrak{f}^*$ of arity $n$, 
called the {\sl function name} of $\mathfrak{f}$. For each pair $\mathfrak{p} =
(\mathfrak{p}_h, \mathfrak{p}_t)$ of subsets of 
$\mathcal{U}^n$ such that $\mathfrak{p}_h \subseteq \mathfrak{p}_t$, we 
introduce an $n$-ary predicate constant $\mathfrak{p}^*$, called the 
{\sl predicate name} of $(\mathfrak{p}_h,\mathfrak{p}_t)$. By $\Sigma^*$ we 
denote the signature obtained by adding all function and predicate names to 
$\Sigma$, and by $\mathcal{U}^*$ we denote the Herbrand universe of $\Sigma^*$.  
Then for each term $\alpha \in \mathcal{U}^*$, we define the term $\w\alpha \in 
\mathcal{U}$ recursively as follows:
\begin{itemize}
\item if $\alpha$ is an object constant from $\mathcal{U}$ then $\w \alpha$ is 
$\alpha$;
\item if $\alpha$ is of the form $f(\alpha_1, \dots, \alpha_n)$ where $f$ is a
function constant from $\Sigma$, then $\w \alpha$ is $f(\w{\alpha_1}, \dots, 
\w{\alpha_n})$;
\item if $\alpha$ is of the form $\mathfrak{f}^*(\alpha_1, \dots, \alpha_n)$ 
where $\mathfrak{f}^*$ is a function name, then $\w \alpha$ is the element of
$\mathcal{U}$ obtained by applying $\mathfrak{f}$ to 
$\langle\w{\alpha_1}, \dots, \w{\alpha_n} \rangle$.
\end{itemize} 

The satisfaction relation between an Herbrand HT-interpretation $J = \langle J^h, 
J^t \rangle$, 
a world $w$, and a closed second-order formula $F$
over $\Sigma$ is defined  recursively, as  follows: 
\begin{enumerate}[(i)]
\item $J,w \not \models \bot$.
\item $J,w \models \alpha_1 = \alpha_2$ if $\w {\alpha_1}$ is $\w 
{\alpha_2}$. 
\item $J,w \models P(\alpha_1, \dots, \alpha_n)$ if $P(\w{\alpha_1}, 
\dots, \w {\alpha_n}) \in J^w$.
\item $J,w \models \mathfrak{p}^*(\alpha_1, \dots, \alpha_n)$ if 
$\langle \w{\alpha_1}, \dots, \w {\alpha_n} \rangle \in \mathfrak{p}_w$. 
\item $J,w \models F \land G$ if $J,w \models F$ and $J,w \models G$; similarly 
for $\lor$. 
\item $J,w\models F\rar G$ if for every world $w'$ such that $w \leq w'$, 
 $J,w'\not\models F$ or $J,w'\models G$.
\item $J,w \models \forall v F$, where $v$ is an object variable, if for each ground term $\alpha$ over~$\Sigma$, 
$J,w \models F^v_{\alpha}$; similarly for~$\exists$.  
\item $J,w \models \forall v F$, where $v$ is a function variable, if for 
each function name $\mathfrak{f}^*$ of the same arity as $v$, $J,w \models 
F^v_{\mathfrak{f}^*}$; similarly for~$\exists$.\footnote{The notation for
substituting
a function name for a function variable is the same as that of substituting
a term for an object variable; similarly for
predicate names and predicate variables.}
\item $J,w \models \forall v F$, where $v$ is a predicate variable, if 
for each predicate name $\mathfrak{p}^*$ of the same arity as $v$, 
$J,w \models F^v_{\mathfrak{p}^*}$; similarly for~$\exists$.  
\end{enumerate}
A closed second-order formula~$F$ over~$\Sigma^*$
is {\sl HHT-valid}  if $J, h \models F$ for every
Herbrand HT-interpretation~$J$.

\noindent {\bf Soundness Lemma. } {\sl 
\begin{enumerate}
\item[(a)] If a second-order formula $F$ over $\Sigma^*$
is provable in 
$\mathbf{HHT}^2$ then the universal closure of $F$ is HHT-valid.
\item[(b)] For any first-order signature $\Sigma$ containing finitely many 
function constants, 
if a second-order formula $F$ over $\Sigma^*$ is provable in 
$\mathbf{HHT}^2 + $ DCA then the universal closure of $F$ is HHT-valid. 
\end{enumerate}
}

The lemma is proved by induction on the derivation of $F$.

\medskip \noindent {\bf Lifting Lemma.} {\sl Let $I$ be an HT-interpretation of a 
propositional signature $\sigma$, $\psi$ be a  substitution from a 
first-order signature $\Sigma$ (possibly containing restrictors) to $\sigma$,
and $J$ be the Herbrand HT-interpretation  
defined by the condition: for every world $w$
$$J,w \models P(\alpha_1, \dots, \alpha_n) \;\text{ iff } \;
I,w \models \psi P(\alpha_1, \dots \alpha_n).$$
Then for any closed first-order formula  $F$ (possibly with restrictors) 
$$J,w \models F \; \text{ iff }  \; I,w \models \psi F.$$}

The lemma is proved by strong induction on the
total number of connectives and quantifiers in $F$.
If $F$ is atomic, then the assertion of the lemma is immediate from 
the definition of $J$.  Here are two of the other cases.

\medskip

\noindent Case $\forall v F$:
\begin{itemize}
  \item[] $J, w \models \forall v F$
  \item[iff] for each ground term $\alpha$, $J, w \models F^v_{\alpha}$
  \item[iff] for each ground term $\alpha$, $I, w \models \psi F^v_{\alpha}$
  \item[iff] $I, w \models \bigwedge_\alpha \psi F^v_{\alpha}$
  \item[iff] $I, w \models \psi \left ( \bigwedge_\alpha F^v_{\alpha} \right ).$
\end{itemize}
 
\noindent Case $\forall (x_1\!:\!R_1,\dots,x_n\!:\!R_n) F$: We need to 
show that 
$$J,w \models \forall (x_1\!:\!R_1,\dots,x_n\!:\!R_n) F$$ iff 
\beq
I, w \models \bigwedge_{\alpha_1,\dots,\alpha_n :
        \ \psi R_1(\alpha_1) = \dots = \psi R_n(\alpha_n) = \top}
        \psi F^{x_1,\cdots,x_n}_{\alpha_1,\cdots,\alpha_n}. 
\eeq{2sho}
Indeed,
\begin{itemize}
  \item[] $J, w \models \forall (x_1\!:\!R_1,\dots,x_n\!:\!R_n) F$
  \item[iff] $J, w \models \forall x_1, \dots, x_n (R_1(x_1) \land \dots \land 
R_n(x_n) \rar F)$
  \item[iff] $J, w' \models F^{x_1,\cdots,x_n}_{\alpha_1,\cdots,\alpha_n}$ 
 in every world $w' \geq w$
    and for each tuple of ground terms $\alpha_1, \dots, \alpha_n$
    such that $J, w' \models R_1(\alpha_1) \land \dots \land 
R_n(\alpha_n) $  
  \item[iff] $I, w' \models \psi F^{x_1,\cdots,x_n}_{\alpha_1,\cdots,\alpha_n} $ 
in every world $w' \geq w$
    and for each tuple of ground terms $\alpha_1, \dots, \alpha_n$ 
   such that $I, w'  \models \psi R_1(\alpha_1)  \land \dots \land 
\psi R_n(\alpha_n)  $ 
  \item[iff] $I, w' \models \psi F^{x_1,\cdots,x_n}_{\alpha_1,\cdots,\alpha_n} $
 in every world $w' \geq w$
    and for each tuple of ground terms $\alpha_1, \dots, \alpha_n$
   such that $ \psi R_1(\alpha_1) = \dots = \psi R_n(\alpha_n) = \top$
  \item[iff] in every world $w' \geq w$,\\ 
  $ I, w' \models \bigwedge_{\alpha_1,\dots,\alpha_n :
        \ \psi R_1(\alpha_1) = \dots = \psi R_n(\alpha_n) = \top}
        \psi F^{x_1,\cdots,x_n}_{\alpha_1,\cdots,\alpha_n}. $
\end{itemize}
The condition above is equivalent to (\ref{2sho}) by the monotonicity property
of the satisfaction relation in the logic of here-and-there.

The main theorem is immediate from the two lemmas stated above.

\section{Conclusion}\label{sec:conc}

In this paper we defined when an infinitary propositional formula is an
instance of a first-order formula.  The provability of first-order formulas
in some extensions of intuitionistic logic implies that all instances of
these formulas are HT-valid.  Theorems of this kind
can be used for establishing the strong equivalence of logic programs
that use local variables ranging over infinite domains.\footnote{If variables 
range over a fixed finite domain then strong equivalence is decidable 
but co-NEXPTIME-complete \cite[Theorem 16]{eit05b}.} 

If an infinite conjunction is an instance of a first-order formula then
it is syntactically uniform, in the sense that all its conjunctive terms are
all of the same kind---either each of them is an atom, or each is an
implication, and so forth.  The same can be said about infinite disjunctions.
This fact points to a limitation on the applicability of the method of
proving HT-validity described in this paper.  For instance, formulas of
the form
\beq
(\neg\neg F_1\lor \neg F_1)\land(F_2\rar F_2)\land
(\neg\neg F_3\lor \neg F_3)\land(F_4\rar F_4)\land\cdots
\eeq{bad}
are HT-valid, but they are not instances of any first-order formula provable
in the deductive systems discussed above.  Indeed, if~(\ref{bad}) is an
instance of a first-order formula~$F$ then~$F$ is either an atom such
that its predicate symbol is not a restrictor of~$F$, or such an atom
preceded by a universally quantified generalized variable.  Such first-order
formulas are not provable.  But it is clear that~(\ref{bad}) can be
tranformed into an instance of a theorem of~$\mathbf{HHT}$ by rewriting
it as a conjunction of two infinite conjunctions:
$$((\neg\neg F_1\lor \neg F_1)\land(\neg\neg F_3\lor \neg F_3)\land\cdots)
\land
((F_2\rar F_2)\land(F_4\rar F_4)\land\cdots).$$
In this sense, the syntactic uniformity of instances of first-order formulas
is not a significant limitation.

\section*{Acknowledgements}

Many thanks to Yuliya Lierler and the anonymous referees for useful comments 
and interesting questions.
The first two authors were 
partially supported by the National Science Foundation under 
Grant IIS-1422455. 

\bibliographystyle{acmtrans}
\bibliography{bib}

\begin{thebibliography}{}

\bibitem[\protect\citeauthoryear{Clark}{Clark}{1978}]{cla78}
{\sc Clark, K.} 1978.
\newblock Negation as failure.
\newblock In {\em Logic and Data Bases}, {H.~Gallaire} {and} {J.~Minker}, Eds.
  Plenum Press, New York, 293--322.

\bibitem[\protect\citeauthoryear{Eiter, Fink, Tomits, and Woltran}{Eiter
  et~al\mbox{.}}{2005}]{eit05b}
{\sc Eiter, T.}, {\sc Fink, M.}, {\sc Tomits, H.}, {\sc and} {\sc Woltran, S.}
  2005.
\newblock Strong and uniform equivalence in answer-set programming:
  Characterizations and complexity results for the non-ground case.
\newblock In {\em Proceedings of AAAI Conference on Artificial Intelligence
  ({AAAI})}. 695--700.

\bibitem[\protect\citeauthoryear{Gebser, Harrison, Kaminski, Lifschitz, and
  Schaub}{Gebser et~al\mbox{.}}{2015}]{geb15}
{\sc Gebser, M.}, {\sc Harrison, A.}, {\sc Kaminski, R.}, {\sc Lifschitz, V.},
  {\sc and} {\sc Schaub, T.} 2015.
\newblock Abstract {G}ringo.
\newblock {\em Theory and Practice of Logic Programming\/}~{\em 15}, 449--463.

\bibitem[\protect\citeauthoryear{Harrison, Lifschitz, Pearce, and
  Valverde}{Harrison et~al\mbox{.}}{2015}]{har15a}
{\sc Harrison, A.}, {\sc Lifschitz, V.}, {\sc Pearce, D.}, {\sc and} {\sc
  Valverde, A.} 2015.
\newblock Infinitary equilibrium logic and strong equivalence.
\newblock In {\em Proceedings of International Conference on Logic Programming
  and Nonmonotonic Reasoning ({LPNMR})}. 398--410.

\bibitem[\protect\citeauthoryear{Harrison, Lifschitz, and
  Truszczynski}{Harrison et~al\mbox{.}}{2015}]{har13c}
{\sc Harrison, A.}, {\sc Lifschitz, V.}, {\sc and} {\sc Truszczynski, M.} 2015.
\newblock On equivalence of infinitary formulas under the stable model
  semantics.
\newblock {\em Theory and Practice of Logic Programming\/}~{\em 15,\/}~1,
  18--34.

\bibitem[\protect\citeauthoryear{Hosoi}{Hosoi}{1966}]{hos66}
{\sc Hosoi, T.} 1966.
\newblock The axiomatization of the intermediate propositional systems~${S}_n$
  of {G}{\"o}del.
\newblock {\em Journal of the Faculty of Science of the University of
  Tokyo\/}~{\em 13}, 183--187.

\bibitem[\protect\citeauthoryear{Lifschitz, Morgenstern, and
  Plaisted}{Lifschitz et~al\mbox{.}}{2008}]{lif08b}
{\sc Lifschitz, V.}, {\sc Morgenstern, L.}, {\sc and} {\sc Plaisted, D.} 2008.
\newblock Knowledge representation and classical logic.
\newblock In {\em Handbook of Knowledge Representation}, {F.~van Harmelen},
  {V.~Lifschitz}, {and} {B.~Porter}, Eds. Elsevier, 3--88.

\bibitem[\protect\citeauthoryear{Lifschitz, Pearce, and Valverde}{Lifschitz
  et~al\mbox{.}}{2007}]{lif07a}
{\sc Lifschitz, V.}, {\sc Pearce, D.}, {\sc and} {\sc Valverde, A.} 2007.
\newblock A characterization of strong equivalence for logic programs with
  variables.
\newblock In {\em Procedings of International Conference on Logic Programming
  and Nonmonotonic Reasoning ({LPNMR})}. 188--200.

\bibitem[\protect\citeauthoryear{Mints}{Mints}{2000}]{min00}
{\sc Mints, G.} 2000.
\newblock {\em A Short Introduction to Intuitionistic Logic}.
\newblock Kluwer.

\bibitem[\protect\citeauthoryear{Truszczynski}{Truszczynski}{2012}]{tru12}
{\sc Truszczynski, M.} 2012.
\newblock Connecting first-order {ASP} and the logic {FO(ID)} through reducts.
\newblock In {\em Correct Reasoning: Essays on Logic-Based AI in Honor of
  Vladimir Lifschitz}, {E.~Erdem}, {J.~Lee}, {Y.~Lierler}, {and} {D.~Pearce},
  Eds. Springer, 543--559.

\bibitem[\protect\citeauthoryear{Umezawa}{Umezawa}{1959}]{ume59}
{\sc Umezawa, T.} 1959.
\newblock On intermediate many-valued logics.
\newblock {\em Journal of the Mathematical Society of Japan\/}~{\em 11,\/}~2,
  116--128.

\end{thebibliography}

\end{document}